\documentclass[11pt,letterpaper]{article}
\usepackage{emnlp2016}
\usepackage{times}
\usepackage{latexsym}
\usepackage{amsfonts}
\usepackage{textcomp}
\usepackage{color}
\usepackage{graphicx}
\usepackage{multirow}
\usepackage{url}
\usepackage{amsmath}
\usepackage{fixltx2e}
\usepackage{subfigure}

\newcommand{\reffig}[1]{Figure~\ref{#1}}
\newcommand{\reftbl}[1]{Table~\ref{#1}}
\newcommand{\refsec}[1]{Section~\ref{#1}}

\newcommand{\system}{RSI}
\newcommand{\systemfull}{Relation Schema Induction}

\newcommand{\Real}{\mathbb{R}}
\newcommand{\cnnr}{SICTF}
\newcommand{\cnnrfull}{Schema Induction using Coupled Tensor Factorization}
\pdfoutput=1

\emnlpfinalcopy 
\def\emnlppaperid{509} 

\title{Relation Schema Induction using Tensor Factorization with Side Information}

\author{Madhav Nimishakavi\\
	    Indian Institute of Science\\
	    Bangalore, India\\
	    {\tt \small{madhav@csa.iisc.ernet.in}}
	  \And
	Uday Singh Saini\\
  	Indian Institute of Science\\
  	Bangalore, India\\
  {\tt \small{uday.s.saini@gmail.com} }
  \And
  Partha Talukdar\\\
  Indian Institute of Science\\
	    Bangalore, India\\
	    {\tt \small{ppt@cds.iisc.ac.in}}}

\date{}

\begin{document}

\maketitle

\begin{abstract}
Given a set of documents from a specific domain (e.g., medical research journals), how do we automatically build a Knowledge Graph (KG) for that domain? Automatic identification of relations and their schemas, i.e., type signature of arguments of relations (e.g., \textit{undergo(Patient, Surgery)}), is an important first step towards this goal. We refer to this problem as \textit{Relation Schema Induction (RSI)}. In this paper, we propose \cnnrfull{} (\cnnr{})
, a novel tensor factorization method for relation schema induction. \cnnr{} factorizes Open Information Extraction (OpenIE) triples extracted from a domain corpus along with additional side information in a principled way to induce relation schemas. To the best of our knowledge, this is the first application of tensor factorization for the RSI problem. Through extensive experiments on multiple real-world datasets, we find that \cnnr{} is not only more accurate than state-of-the-art baselines, but also significantly faster (about 14x faster).

\end{abstract}

\section{Introduction}
\label{sec:intro}

\begin{table*}[t]
\begin{scriptsize}
\centering
 \begin{tabular}{|p{3cm}|p{2cm}|p{1.7cm}|p{2.2cm}|p{2cm}|p{2cm}|}
  \hline

   & Target task & Interpretable latent factors? & Can induce relation schema? & Can use NP side info? & Can use relation side info? \\
  \hline
Typed RESCAL \cite{export:226677}  & Embedding & No & No & Yes & No \\ \hline
Universal Schema \cite{singh2015towards}  & Link Prediction & No & No & No & No \\ \hline
KB-LDA \cite{kblda:movshovitzattias-wcohen:2015:ACL}  & Ontology Induction & Yes & Yes & Yes & No \\ \hline
\cnnr{} (this paper)  & Schema Induction & Yes & Yes & Yes & Yes \\ \hline

 \end{tabular}
\caption{\label{tbl:related_work} Comparison among \cnnr{} (this paper) and other related methods. KB-LDA  is the most related prior method which is extensively compared against \cnnr{} in \refsec{sec:expts}}

\end{scriptsize}
\end{table*}

Over the last few years, several techniques to build Knowledge Graphs (KGs) from large unstructured text corpus have been proposed, examples include NELL \cite{NELL-aaai15} and Google Knowledge Vault \cite{dong2014knowledge}. Such KGs consist of millions of entities (e.g., \textit{Oslo, Norway}, etc.), their types (e.g., \textit{isA(Oslo, City)}, \textit{isA(Norway, Country)}), and relationships among them (e.g., \textit{cityLocatedInCountry(Oslo, Norway)}). These KG construction techniques are called ontology-guided as they require as input list of relations, their schemas (i.e., their type signatures, e.g., \textit{cityLocatedInCountry(City, Country)}), and seed instances of each such relation. Listing of such relations and their schemas are usually prepared by human domain experts.

The reliance on domain expertise poses significant challenges when  such ontology-guided KG construction techniques are applied to domains where domain experts are either not  available or are too expensive to employ. Even when such a domain expert may be available for a limited time, she may be able to provide only a partial listing of relations and their schemas relevant to that particular domain. Moreover, this expert-mediated model is not scalable when new data in the domain becomes available, bringing with it potential new relations of interest. In order to overcome these challenges, we need automatic techniques which can discover relations and their schemas from unstructured text data itself, without requiring extensive human input. We refer to this problem as \emph{\systemfull{} (\system{})}.



In contrast to ontology-guided KG construction techniques mentioned above, Open Information Extraction (OpenIE) techniques \cite{etzioni2011open} aim to extract surface-level triples from unstructured text. Such OpenIE triples may provide a suitable starting point for the \system{} problem. 
In fact, KB-LDA, a topic modeling-based method for inducing an ontology from SVO (Subject-Verb-Object) triples was recently proposed in \cite{kblda:movshovitzattias-wcohen:2015:ACL}. We note that ontology induction \cite{velardi2013ontolearn} is a more general problem than \system{}, as we are primarily interested in identifying categories and relations from a domain corpus, and not necessarily any hierarchy over them. Nonetheless, KB-LDA maybe used for the \system{} problem and we use it as a representative of the state-of-the-art of this area.  

Instead of a topic modeling approach, we take a tensor factorization-based approach for \system{} in this paper. Tensors are a higher order generalization of matrices and they provide a natural way to represent OpenIE triples. 
Applying tensor factorization methods over OpenIE triples to identify relation schemas is a natural approach, but one that has not been explored so far. Also, a tensor factorization-based approach presents a flexible and principled way to incorporate various types of side information. Moreover, as we shall see in \refsec{sec:expts}, compared to state-of-the-art baselines such as KB-LDA, tensor factorization-based approach results in better and faster solution for the \system{} problem. In this paper, we make the following contributions:
	\begin{itemize}
		\item We present \cnnrfull{} (\cnnr{})
, a novel and principled tensor factorization method which jointly factorizes a tensor constructed out of OpenIE triples extracted from a domain corpus, along with various types of additional side information for relation schema induction.
		\item We compare \cnnr{} against state-of-the-art baseline on various real-world datasets from diverse domains. We observe that \cnnr{} is not only significantly more accurate than such baselines, but also much faster. For example, \cnnr{} achieves 14x speedup over KB-LDA \cite{kblda:movshovitzattias-wcohen:2015:ACL}.
		\item  We have made the data and code available \footnote{https://github.com/malllabiisc/sictf}. 
	\end{itemize}

\begin{figure*}[t]
\begin{small}
 \centering
 \includegraphics[scale=0.37]{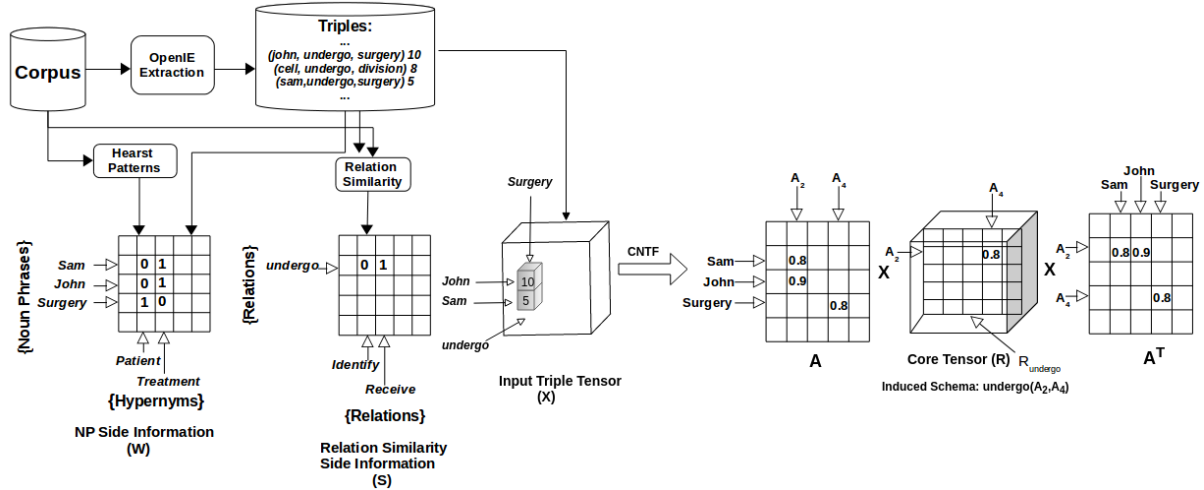}
 \caption{\label{fig:cntf_arch}Relation Schema Induction (RSI) by \cnnr{}, the proposed method. First, a tensor ($X$) is constructed to represent OpenIE triples extracted from a domain corpus. Noun phrase side information in the form of (noun phrase, hypernym), and relation-relation similarity side information are separately calculated and stored in two separate matrices ($W$ and $S$, respectively). \cnnr{} then performs coupled factorization of the tensor and the two side information matrices to identify relation schemas which are stored in the core tensor ($R$) in the output. Please see \refsec{sec:method} for details.}
 \label{fig1}
 \end{small}
\end{figure*}

\section{Related Work}
\label{sec:related}

%
%
%

{\bf Schema Induction}: 
Properties of \cnnr{} and other related methods are summarized in \reftbl{tbl:related_work}\footnote{Please note that not all methods mentioned in the table are directly comparable with \cnnr{}, the table only illustrates the differences.
KB-LDA is the only method which is directly comparable.}. A method for inducing (binary) relations and the categories they connect was proposed by \cite{Mohamed11discoveringrelations}. However, in that work, categories and their instances were known a-priori. In contrast, in case of \cnnr{}, both categories and relations are to be induced.
A method for event schema induction, the task of learning high-level representations of complex events and their entity roles from unlabeled text, was proposed in \cite{Chambers13}. This gives the schemas of slots per event, but our goal is to find schemas of relations.
\cite{chen-etal-2013} and \cite{ChenWGR15} deal with the problem of finding semantic slots for unsupervised spoken language understanding, but we are interested in finding schemas of  relations relevant for a given domain. \\
Methods for link prediction in the Universal Schema setting using matrix and a combination of matrix and tensor factorization are proposed in \cite{RiedelYMM13} and \cite{singh2015towards}, respectively. Instead of link prediction where relation schemas are assumed to be given, \cnnr{} focuses on discovering such relation schemas. Moreover, in contrast to such methods which assume access to existing KGs, the setting in this paper is unsupervised.

{\bf Tensor Factorization}: 
Due to their flexibility of representation and effectiveness, tensor factorization methods have seen increased application in Knowledge Graph (KG) related problems over the last few years. Methods for decomposing ontological KGs such as YAGO \cite{suchanek2007yago} were proposed in \cite{Nickel:2012:FYS:2187836.2187874,chang2014typed,export:226677}. In these cases, relation  schemas are known in advance, while we are interested in inducing such relation schemas from unstructured text. A PARAFAC \cite{harshman1970fpp} based method for jointly factorizing a matrix and tensor for data fusion was proposed in \cite{Acar2013}. In such cases, the matrix is used to provide auxiliary information \cite{usingaux,Erdos:2013:DFB:2505515.2507846}. 
Similar PARAFAC-based ideas are explored in Rubik \cite{conf/kdd/WangCGDKCMS15} 
to factorize structured electronic health records. In contrast to such structured data sources, \cnnr{} aims at inducing relation schemas from unstructured text data. 
Propstore, a tensor-based model for distributional semantics, a problem different from \system{}, was presented in \cite{Goyal_astructured}. Even though coupled factorization of tensor and matrices constructed out of unstructured text corpus provide a natural and plausible  approach for the \system{} problem, they have not yet been explored -- we fill this gap in this paper.

{\bf Ontology Induction}: 
\systemfull{}  can be considered a sub problem of Ontology Induction \cite{velardi2013ontolearn}. 
Instead of building a full-fledged hierarchy over categories and relations as in ontology induction, we are particularly interested in finding relations and their schemas from unstructured text corpus. We consider KB-LDA\footnote{In this paper, whenever we refer to KB-LDA, we only refer to the part of it that learns relations from unstructured data.} \cite{kblda:movshovitzattias-wcohen:2015:ACL}, a topic-modeling based approach for ontology induction, as a representative of this area. Among all prior work, KB-LDA is most related to \cnnr{}. While both KB-LDA and \cnnr{} make use of noun phrase side information, \cnnr{} is also able to exploit relational side information in a principled manner.  
  In \refsec{sec:expts}, through experiments on multiple real-world datasets, we observe that \cnnr{} is not only more accurate than KB-LDA but also significantly faster with a speedup of 14x. 

  A method for canonicalizing noun and relation phrases in OpenIE triples was recently proposed in \cite{galarraga2014canonicalizing}. The main focus of this approach is to cluster lexical variants of a \emph{single} entity or relation. This is not directly relevant for \system{}, as we are interested in grouping \emph{multiple} entities of the same type into one cluster, and use that to induce relation schema. 
\section{Our Approach: \cnnrfull{} (\cnnr{})}
\label{sec:method}


\subsection{Overview}
\label{sec:overview}

\cnnr{} poses the relation schema induction problem as a coupled factorization of a tensor along with matrices containing relevant side information. 
Overall architecture of the \cnnr{} system is presented in \reffig{fig:cntf_arch}. First, a tensor $X \in \Real_{+}^{n \times n \times m}$ is constructed to store OpenIE triples and their scores extracted from the text corpus\footnote{$\Real_{+}$ is the set of non-negative reals.}. Here, $n$ and $m$ represent the number of NPs and relation phrases, respectively.
Following \cite{kblda:movshovitzattias-wcohen:2015:ACL}, \cnnr{} makes use of noun phrase (NP)  side information in the form of (noun phrase, hypernym).
Additionally, \cnnr{} also exploits relation-relation similarity side information. These two side information are stored in matrices $W \in \{0,1\}^{n \times h}$ and $S \in \{0,1\}^{m \times m}$, where $h$ is the number of hypernyms extracted from the corpus. \cnnr{} then performs collective non-negative factorization over $X$, $W$, and $S$ to output  matrix $A \in \Real_{+}^{n \times c}$ and the core tensor $R \in \Real_{+}^{c \times c \times m}$. Each row in $A$ corresponds to an NP, while each column corresponds to an induced category (latent factor). For brevity, we shall refer to the induced category corresponding to the $q^{th}$ column of $A$ as $A_q$. Each entry $A_{pq}$ in the output matrix provides a membership score for NP $p$ in induced category $A_q$. 
Please note that each induced category is represented using the NPs participating in it, with the NPs ranked by their membership scores in the induced category. In \reffig{fig:cntf_arch}, {\small $A_2 = [(John, 0.9), (Sam, 0.8), \ldots]$} is an induced category.

Each slice of the core tensor $R$ is a matrix which corresponds to a specific relation, e.g., the matrix $R_{undergo}$ highlighted in \reffig{fig:cntf_arch} corresponds to the relation \textit{undergo}. 
Each cell in this matrix corresponds to an induced schema connecting two induced categories (two columns of the $A$ matrix), with the cell value representing model's score of the induced schema. 
For example, in \reffig{fig:cntf_arch}, \textit{undergo($A_2$, $A_4$)} is an induced relation schema with score $0.8$ involving relation \textit{undergo} and induced categories $A_2$ and $A_4$. 

In \refsec{sec:side_info}, we present details of the side information used by \cnnr{}, and then in \refsec{sec:model_details} present details of the optimization problem solved by \cnnr{}.


\subsection{Side Information}
\label{sec:side_info}


%

\begin{table}[t]
\begin{scriptsize}
\centering
 \begin{tabular}{|p{7cm}|}
  \hline
   MEDLINE \\
  \hline
  (hypertension, disease), (hypertension, state), (hypertension, disorder) ,
  (neutrophil, blood element), (neutrophil, effector cell), (neutrophil, cell type) \\
  \hline
  \hline
   StackOverflow \\
   \hline
   (image, resource), (image, content), (image, file), (perl, language), (perl, script), (perl, programs) \\
  \hline
 \end{tabular}
\caption{\label{tbl:np_cats}Noun Phrase (NP) side information in the form of (Noun Phrase, Hypernym) pairs extracted using Hearst patterns from two different datasets. Please see \refsec{sec:side_info} for details.}
\end{scriptsize}
\end{table}

\begin{table}[t]
\begin{scriptsize}
\centering
 \begin{tabular}{|p{3.5cm}|p{3.5cm}|}
  \hline

   \multicolumn{1}{|l|}{MEDLINE} & \multicolumn{1}{|l|}{StackOverflow} \\
  \hline
(evaluate, analyze), (evaluate, examine), (indicate, confirm), (indicate, suggest) & (provides, confirms), (provides, offers), (allows, lets), (allows, enables) 
\\ \hline
 \end{tabular}
\caption{\label{tbl:sim_verbs}Examples of relation similarity side information in the form of automatically identified similar relation pairs. Please see \refsec{sec:side_info} for details.}
\end{scriptsize}
\end{table}

\begin{itemize}
	\item {\bf Noun Phrase Side Information}: 
	Through this type of side information, we would like to capture type information of as many noun phrases (NPs) as possible. 
	We apply Hearst patterns \cite{Hearst1992}, e.g., \textit{"$<$Hypernym$>$ such as $<$NP$>$"}, over the corpus to extract such \textit{(NP, Hypernym)} pairs. Please note that neither hypernyms nor NPs are pre-specified, and they are all extracted from the data by the patterns. Examples of a few such pairs extracted from two different datasets are shown in \reftbl{tbl:np_cats}. These extracted tuples are stored in a matrix $W_{n \times h}$ whose rows correspond to NPs and columns correspond to extracted hypernyms. We define,		
		\[
		\begin{scriptsize}
		  W_{ij} = \begin{cases}
            1, & \text{if NP\textsubscript{i} belongs to Hypernym\textsubscript{j}} \\
            0, & \text{otherwise}
           \end{cases} .
 		\end{scriptsize}
 		\]
 		Please note that we don't expect $W$ to be a fully specified matrix, i.e., we don't assume that we know all possible hypernyms for a given NP. 
	\item {\bf Relation Side Information}: In addition to the side information involving NPs, we would also like to take prior knowledge about textual relations into account during factorization. 
	For example, if we know two relations to be similar to one another, then we also expect their induced schemas to be similar as well. 
	Consider the following sentences \textit{"Mary purchased a stuffed animal toy.``} and \textit{"Janet bought a toy car for her son.''}. From these we can say that both relations \textit{purchase} and \textit{buy} have the schema 
	(Person, Item). Even if one of these relations is more abundant than the other in the corpus, we still want to learn similar schemata for both the relations.
	As mentioned before, $S \in \Real_{+}^{m \times m}$ is the relation similarity matrix, where $m$ is the number of textual relations. We define,
		\[
			\begin{scriptsize}
		  S_{ij} = \begin{cases}
            1, & \text{if Similarity(Rel\textsubscript{i}, Rel\textsubscript{j}) $\ge \gamma$} \\
            0, & \text{otherwise}
           \end{cases}
	 		\end{scriptsize}           
 		\]
	where $\gamma$ is a threshold\footnote{For the experiments in this paper, we set $\gamma = 0.7$, a relatively high value, to focus on highly similar relations and thereby justifying the binary $S$ matrix.}.
	For the experiments in this paper, we use cosine similarity over word2vec \cite{NIPS2013_5021} vector representations of the relational phrases. Examples of a few similar relation pairs are shown in \reftbl{tbl:sim_verbs}.
\end{itemize}

\subsection{\cnnr{} Model Details}
\label{sec:model_details}


\cnnr{} performs coupled non-negative factorization of the input triple tensor $X_{n \times n \times m}$ along with the two side information matrices $W_{n \times h}$ and $S_{m \times m}$ by solving the following  optimization problem.

\vspace{-0.4cm}
\begin{small}
\begin{equation}
 \min\limits_{A, V, R} \sum_{k = 1}^{m} f(X_{k},  A, R_{k}) + f_{np}(W, A, V) + f_{rel}(S, R) 
\end{equation}

where,
\vspace{-0.2cm}
\begin{align*}
 f(X_{k}, A,R_{k}) &= \parallel X_{:,:,k} - AR_{:,:,k}A^T\parallel_{F}^2 + \lambda_R  \parallel R_{:,:,k} \parallel_F^2 \nonumber \\
 f_{np}(W, A, V) &= \lambda_{np} \parallel W-AV \parallel_{F}^2 + \lambda_A \parallel A \parallel_F^2 \nonumber \\
 	&~~~~~~~~~~~+ \lambda_V \parallel V \parallel_F^2 \nonumber \\
 f_{rel}(S, R) &= \lambda_{rel} \sum\limits_{i=1}^{m} \sum\limits_{j=1}^{m} S_{ij} \parallel R_{:,:,i} - R_{:,:,j} \parallel_{F}^2 \nonumber \\ 
 A_{i,j} \ge 0, & V_{j,r} \ge 0, R_{p,q,k} \ge 0 \tag{non negative} \label{eqn:non-neg-constr} \\ 
 & \forall~1 \le i \le n, 1 \le r \le h, \\ 
 & 1 \le j,p,q\le c, 1 \le k \le m \nonumber
\end{align*}
\end{small}
%
In the objective above, the first term $f(X_k, A,R_k)$ minimizes reconstruction error for the $k^{th}$ relation, with additional regularization on the $R_{:,:,k}$ matrix\footnote{For brevity, we also refer to $R_{:,:,k}$ as $R_{k}$, and similarly $X_{:,:,k}$ as $X_{k}$}. The second term, $f_{np}(W, A, V)$, factorizes the NP side information matrix $W_{n \times h}$ into two matrices $A_{n \times c}$ and $V_{c \times h}$, where $c$ is the number of induced categories. We also enforce $A$ to be non-negative. Typically, we require $c \ll h$ to get a lower dimensional embedding of each NP (rows of $A$).
Finally, the third term $f_{rel}(S, R)$ enforces the requirement that two similar relations as given by the matrix $S$ should have similar signatures (given by the corresponding $R$ matrix). Additionally, we require  $V$ and $R$ to be non-negative, as marked by the (non-negative) constraints. In this objective, $\lambda_R$, $\lambda_{np}$, $\lambda_{A}$, $\lambda_{V}$, and $\lambda_{rel}$ are all hyper-parameters.

%
%

 We derive non-negative multiplicative updates for $A$, $R_k$ and $V$ following the rules proposed in \cite{Lee00algorithmsfor}, which has the following general form: 
 \vspace{-5mm}
 
 \begin{equation*}
  \theta_i = \theta_i \left(\frac{\frac{\partial C(\theta)^-}{\partial \theta_i}}{\frac{\partial C(\theta)^+}{\partial \theta_i}}\right)^{\alpha}
 \end{equation*}

 Here $C(\theta)$ represents the cost function of the non-negative variables $\theta$ and $\frac{\partial C(\theta)^-}{\partial \theta_i}$ and $\frac{\partial C(\theta)^-}{\partial \theta_i}$ are the negative and positive parts of the 
 derivative of $C(\theta)$ \cite{IMM2008-04658}. \cite{Lee00algorithmsfor} proved that for $\alpha = 1$, the cost function $C(\theta)$ monotonically decreases with the multiplicative updates \footnote{We also use $\alpha =1$.}. $C(\theta)$ for \cnnr{} is given in equation (1). The above procedure will give the following updates:

\begin{small}
\begin{eqnarray*}
  A &\leftarrow& A * \frac{\sum\limits_k(X_kAR_{k}^T + X_k^TAR_k) + \lambda_{np}WV^T}{A (\tilde{B}+\lambda_AI + \lambda_{np} VV^T)} \\
 \tilde{B} &=& \sum\limits_k (R_kA^TAR_k^T + R_k^TA^TAR_k) \\
  R_k &\leftarrow& R_k * \frac{A^{T}X_{k}A + 2~ \lambda_{rel}~\sum\limits_{j=1}^{m}R_jS_{kj}}{A^TAR_kA^TA + \tilde{D}} \\
 \tilde{D} &=& 2~\lambda_{rel}~R_k \sum\limits_{j=1}^{m}S_{kj} + \lambda_R R_k \\
 V &\leftarrow& V * \frac{\lambda_{np}A^{T}W}{\lambda_{np}A^{T}AV + \lambda_VV} \\
\end{eqnarray*}
\end{small}
In the equations above, $*$ is the Hadamard or element-wise product\footnote{$(A * B)_{i,j} = A_{i,j} \times B_{i,j}$}. In all our experiments, we find the iterative updates above to converge in about 10-20 iterations. 

\section{Experiments}
\label{sec:expts}

 \begin{table}[t]
 \begin{scriptsize}
\begin{center}
 \begin{tabular}{ |p{2.5cm}|c|c|  }
 \hline
 \textbf{Dataset} & \textbf{\# Docs} &\textbf{\# Triples}  \\
 \hline
	 MEDLINE   &  50,216 &  2,499\\
	 \hline
	 StackOverflow & 5.5m    &  37,439  \\
 \hline
\end{tabular}
\caption{\label{tbl:datasets}Datasets used in the experiments.}
\end{center}
\end{scriptsize}
\end{table}

\begin{table*}[t]
\begin{scriptsize}
 \centering
 \begin{tabular}{|p{3cm}|p{9cm}|p{3cm}|}
 \hline
 Relation Schema & Top 3 NPs in Induced Categories which were presented to annotators & Annotator Judgment \\
  \hline
\multicolumn{3}{|c|}{StackOveflow} \\
 \hline
\multirow{2}{*}{\textit{clicks}$(A_0, A_1)$}  & $A_0$: \textit{users, client, person} & \multirow{2}{*}{valid}  \\
   & $A_1$: \textit{link, image, item} & \\
  \hline
  \multirow{2}{*}{\textit{refreshes}$(A_{19},A_{13})$}  & $A_{19}$: \textit{browser, window, tab}  & \multirow{2}{*}{valid}\\ 
   & $A_{13}$: \textit{page, activity, app} & \\  
   \hline
   \multirow{2}{*}{\textit{can\_parse}$(A_{41},A_{17})$}  & $A_{41}$: \textit{access, permission, ability} & \multirow{2}{*}{invalid}\\ 
   & $A_{17}$: \textit{image file, header file, zip file} & \\  
  
  \hline
  \hline
  \multicolumn{3}{|c|}{MEDLINE} \\
  \hline
  \multirow{2}{*}{\textit{suffer\_from}$(A_{38},A_{40})$}  & $A_{38}$: \textit{patient, first patient, anesthetized patient}  & \multirow{2}{*}{valid} \\ 
   & $A_{40}$: \textit{viral disease, renal disease, von recklin	ghausen's disease} &\\ \hline
  
  \multirow{2}{*}{\textit{have\_undergo}$(A_{3},A_{37})$}  & $A_3$: \textit{fifth patient, third patient, sixth patient} & \multirow{2}{*}{valid}\\ 
   & $A_{37}$: \textit{initial liver biopsy, gun biopsy, lymph node biopsy} &\\ \hline
  
  \multirow{2}{*}{\textit{have\_discontinue}$(A_{41},A_{20})$}  & $A_{41}$: \textit{patient, group, no patient} & \multirow{2}{*}{invalid}\\ 
   & $A_{20}$: \textit{endemic area, this area, fiber area} &\\ \hline
  
  \hline
 
 \end{tabular}
 
 \caption{\label{tbl:predicate_examples}Examples of relation schemas induced by \cnnr{} from the StackOverflow and MEDLINE datasets. Top NPs from each of the induced categories, along with human judgment of the induced schema are also shown. See \refsec{sec:main_result} for more details.}
\end{scriptsize}
 \end{table*}


In this section, we evaluate performance of different methods on the  Relation Schema Induction (RSI) task. Specifically, we address the following questions.
\begin{itemize}
 \item Which method is most effective on the RSI task? (\refsec{sec:main_result})
 \item How important are the additional side information for RSI? (\refsec{sec:ablation_result})
 \item What is the importance of non-negativity in RSI with tensor factorization? (\refsec{sec:nn_result})
\end{itemize}

\subsection{Experimental Setup}
\label{sec:expt_setup}

{\bf Datasets}: We used two datasets for the experiments in this paper, they are summarized in \reftbl{tbl:datasets}. 
 For MEDLINE dataset, we used Stanford CoreNLP \cite{manning-EtAl:2014:P14-5} for coreference resolution and Open IE v4.0\footnote{Open IE v4.0: \tiny{{\tt http://knowitall.github.io/openie/}}} for triple extraction. 
 Triples with Noun Phrases that have Hypernym information were retained. We obtained the StackOverflow triples directly from the authors of \cite{kblda:movshovitzattias-wcohen:2015:ACL}, which were also prepared using a very similar process. In both datasets, we use corpus frequency of triples for constructing the tensor.

{\bf Side Information}: Seven Hearst patterns such as  \textit{"$<$hypernym$>$ such as $<$NP$>$", "$<$NP$>$ or other $<$hypernym$>$"} etc., given in \cite{Hearst1992} were used to extract NP side information from the MEDLINE documents. NP side information for the StackOverflow dataset was obtained from the authors of \cite{kblda:movshovitzattias-wcohen:2015:ACL}.

As described in \refsec{sec:method}, word2vec embeddings of the relation phrases were used to extract relation-similarity based side-information. This was done for both datasets. Cosine similarity threshold of $\gamma = 0.7$ was used for the experiments in the paper. 

Samples of side information used in the experiments are shown in \reftbl{tbl:np_cats} and \reftbl{tbl:sim_verbs}. A total of 2067 unique NP-hypernym pairs were extracted from MEDLINE data and 16,639 were from StackOverflow data. 25 unique pairs of relation phrases out of 1172 were found to be similar in MEDLINE data, whereas  280 unique pairs of relation phrases out of approximately 3200 were found similar in StackOverflow data.

Hyperparameters were tuned using grid search and the set which gives minimum reconstruction error for both $X$ and $W$ was chosen. We set $\lambda_{np} = \lambda_{rel} = 100$ for StackOverflow, and $\lambda_{np} = 0.05$ and $\lambda_{rel} = 0.001$ for Medline and we use $c = 50$ for our experiments. Please note that our setting is unsupervised, and hence there is no separate train, dev and test sets. 



 \begin{figure*}[t]
 
 \centering
 \subfigure[\label{fig:predicate_induction}] {\includegraphics[scale=0.4]{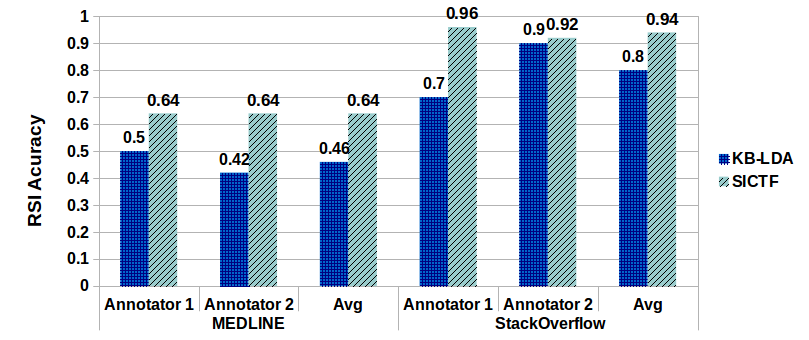}}\quad
 \subfigure[\label{fig:runtime_compare}] {\includegraphics[scale=0.4]{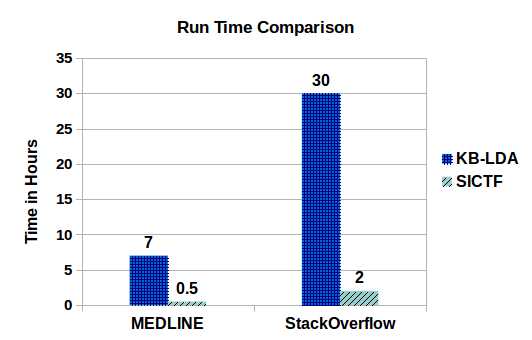}}
 \vspace{-0.5cm}
\caption{\label{fig:main_result}(a) Relation Schema Induction (RSI) accuracies of different methods on the two datasets. \cnnr{}, our proposed method, significantly outperforms state-of-the-art method KBLDA. This is the main result of the paper. Results for KB-LDA on StackOveflow are directly taken from the paper. Please see \refsec{sec:main_result} for details. (b) Runtime comparison between KB-LDA and \cnnr{}. We observe that \cnnr{} results in 14x speedup over KB-LDA. Please see \refsec{sec:main_result} (Runtime Comparison) for details.}
 
 \end{figure*}

 \begin{table*}[t]
\centering
 \begin{tabular}{|c|c|c|c|c|c|c|}
  \hline

   \multirow{2}{*}{Ablation}& \multicolumn{3}{|c|}{MEDLINE} &\multicolumn{3}{|c|}{StackOverflow} \\
  \cline{2-7}
  & A1 & A2 & Avg & A1 & A2 & Avg \\ \hline \hline
\cnnr{}  & 0.64&  0.64& \textbf{0.64} & 0.96 & 0.92 &  \textbf{0.94}\\ \hline
\cnnr{} ($\lambda_{rel}$ = 0) & 0.60 &  0.56 & 0.58 & 0.83 & 0.70 & 0.77 \\ \hline
\cnnr{} ($\lambda_{np}$ = 0) &  0.46& 0.40 & 0.43 & 0.89 & 0.90 & 0.90 \\ \hline
\cnnr{} ($\lambda_{rel}$=0, $\lambda_{np}$ = 0)& 0.46& 0.50& 0.48&0.84 &0.33 & 0.59 \\ \hline
\cnnr{} ($\lambda_{rel}$=0, $\lambda_{np}$ = 0, and no non-negativity constraints ) &0.14 & 0.10&0.12 & 0.20& 0.14 &0.17 \\ \hline

 \end{tabular}
\caption{\label{tbl:ablation}RSI accuracy comparison of \cnnr{} with its ablated versions when no relation side information is used ($\lambda_{rel}$ = 0), when no NP side information is used ($\lambda_{np}$ = 0), when no side information of any kind is used ($\lambda_{rel}$ = 0, $\lambda_{np}$ = 0), and 
when additionally there are no non-negative constraints. From this, we observe that additional side information improves performance, validating one of the central thesis of this paper. Please see  \refsec{sec:ablation_result} and \refsec{sec:nn_result} for details.
}
\end{table*}

 \subsection{Evaluation Protocol}
\label{sec:protocol}
In this section, we shall describe how the induced schemas are presented to human annotators and how final accuracies are calculated. 
In factorizations produced by \cnnr{} and other ablated versions of \cnnr{}, we first select a few top relations with best reconstruction score. The schemas induced for each selected relation $k$ is represented by the matrix slice $R_k$ of the core tensor obtained after factorization 
(see \refsec{sec:method}). 
From each such matrix, we identify the indices $(i, j)$ with highest values. 
The indices $i$ and $j$ select columns of the matrix $A$. A few top ranking NPs  from the columns $A_{i}$ and $A_{j}$ along with the relation $k$ are presented to the human annotator, who then evaluates whether the tuple $\mathrm{Relation}_k(A_{i}, A_{j})$  constitutes a valid schema for relation $k$. Examples of a few relation schemas induced by \cnnr{} are presented in \reftbl{tbl:predicate_examples}. 
A human annotator would see the first and second columns of this table and then offer judgment as indicated in the third column of the table. 
All such judgments across all top-reconstructed relations are aggregated to get the final accuracy score. This evaluation protocol was also used in \cite{kblda:movshovitzattias-wcohen:2015:ACL} to measure learned relation accuracy.

All evaluations were blind, i.e., the annotators were not aware of the method that generated the output they were evaluating. Moreover, the annotators are experts in software domain and has high-school level knowledge in medical domain.
Though recall is a desirable statistic to measure, it is very challenging to calculate it in our setting due to the non-availability of relation schema annotated text on large scale.
\subsection{Results}
\subsubsection{Effectiveness of \cnnr{}}
\label{sec:main_result}
Experimental results comparing performance of various methods on the RSI task in the two datasets are presented in \reffig{fig:predicate_induction}. RSI accuracy is calculated based on the evaluation protocol described in \refsec{sec:protocol}. 
Performance number of KB-LDA for StackOveflow dataset is  taken directly from the \cite{kblda:movshovitzattias-wcohen:2015:ACL} paper, we used our implementation of KB-LDA for the MEDLINE dataset. 
 Annotation accuracies from two annotators were averaged to get the final accuracy. From \reffig{fig:predicate_induction}, we observe that \cnnr{} outperforms  KB-LDA on the RSI task. Please note that the inter-annotator agreement for \cnnr{} is  88\% and 97\% for MEDLINE and StackOverflow datasets respectively. This is the main result of the paper. 


In addition to KB-LDA, we also compared \cnnr{} with PARAFAC, a standard tensor factorization method. PARAFAC induced extremely poor and small number of relation schemas, and hence we didn't consider it any further.

{\bf Runtime comparison}: Runtimes of \cnnr{} and KB-LDA over both datasets are compared in \reffig{fig:runtime_compare}. From this figure, we find that \cnnr{} is able to achieve a 14x speedup on average over KB-LDA\footnote{Runtime of KB-LDA over the StackOverflow dataset was obtained from the authors of \cite{kblda:movshovitzattias-wcohen:2015:ACL} through personal communication. Our own implementation also resulted in similar runtime over this dataset.}. In other words,  \cnnr{} is not only able to induce better relation schemas, but also do so at a significantly faster speed.

\subsubsection{Importance of Side Information}
\label{sec:ablation_result}
One of the central hypothesis of our approach is that coupled factorization through additional side information should result in better relation schema induction. In order to evaluate this thesis further, we compare performance of \cnnr{} with its ablated versions: (1) \cnnr{} ($\lambda_{rel}$ = 0), which corresponds to the setting when no relation side information is used, (2) \cnnr{} ($\lambda_{np} = 0$), which corresponds to the setting when no noun phrases side information is used, and (3) \cnnr{} ($\lambda_{rel}$ = 0, $\lambda_{np} = 0$), which corresponds to the setting when no side information of any kind is used. Hyperparameters are separately tuned for the variants of \cnnr{}.
Results are presented in the first four rows of \reftbl{tbl:ablation}. From this, we observe that additional coupling through the side information significantly helps improve \cnnr{} performance. This further validates the central thesis of our paper.

\subsubsection{Importance of Non-Negativity on Relation Schema Induction}
\label{sec:nn_result}

In the last row of \reftbl{tbl:ablation}, we also present an ablated version of \cnnr{} when no side information no non-negativity constraints are used. Comparing the last two rows of this table, we observe that non-negativity constraints over the $A$ matrix and core tensor $R$ result in significant improvement in  performance. 
We note that the last row in \reftbl{tbl:ablation} is equivalent  to RESCAL \cite{Nickel_athree-way} and the fourth row is equivalent to Non-Negative RESCAL \cite{krompass2013non}, two tensor factorization techniques. We also note that none of these tensor factorization techniques have been previously used for the relation schema induction problem.

The reason for this improved performance may be explained by the fact that absence of non-negativity constraint results in an under constrained factorization problem where the model often overgenerates incorrect triples, and then compensates for this overgeneration by using negative latent factor weights. In contrast, imposition of non-negativity constraints restricts the model further forcing it to commit to specific semantics of the latent factors in $A$. This improved interpretability also results in better RSI accuracy as we have seen above. Similar benefits of non-negativity on interpretability have also been observed in matrix factorization \cite{murphy2012learning}.

\section{Conclusion}
\label{sec:conclusion}

Relation Schema Induction (RSI) is an important first step towards building a Knowledge Graph (KG) out of text corpus from a given domain. While human domain experts have traditionally prepared listing of relations and their schemas, this expert-mediated model poses significant challenges in terms of scalability and coverage. In order to overcome these challenges, in this paper, we present \cnnr{}, a novel non-negative coupled tensor matrix factorization method for relation schema induction. \cnnr{} is flexible enough to incorporate various types of side information during factorization. Through extensive experiments on real-world datasets, we find that \cnnr{} is not only more accurate but also significantly faster (about 11.8x speedup) compared to state-of-the-art baselines. As part of future work, we hope to analyze CNTF and its optimization further, assign labels to induced categories, and also apply the model to more domains. We hope to make all code and datasets used in the paper publicly available upon publication of the paper.

\section*{Acknowledgement} 
  Thanks to the members of MALL Lab, IISc who read our drafts and gave valuable feedback and we also thank the reviewers for their constructive reviews.
  This research has been supported in part by Bosch Engineering and Business Solutions and Google.

\bibliography{directed_ont_ext}

\begin{thebibliography}{}

\bibitem[\protect\citename{Acar \bgroup et al.\egroup }2013]{Acar2013}
Evrim Acar, Morten~Arendt Rasmussen, Francesco Savorani, Tormod Næs, and
  Rasmus Bro.
\newblock 2013.
\newblock Understanding data fusion within the framework of coupled matrix and
  tensor factorizations.
\newblock {\em Chemometrics and Intelligent Laboratory Systems},
  129(Complete):53--63.

\bibitem[\protect\citename{Chambers}2013]{Chambers13}
Nathanael Chambers.
\newblock 2013.
\newblock Event schema induction with a probabilistic entity-driven model.
\newblock In {\em EMNLP}, pages 1797--1807. ACL.

\bibitem[\protect\citename{Chang \bgroup et al.\egroup }2014a]{export:226677}
Kai-Wei Chang, Wen tau Yih, Bishan Yang, and Christopher Meek.
\newblock 2014a.
\newblock Typed tensor decomposition of knowledge bases for relation
  extraction.
\newblock In {\em Proceedings of the 2014 Conference on Empirical Methods in
  Natural Language Processing}. ACL – Association for Computational
  Linguistics, October.

\bibitem[\protect\citename{Chang \bgroup et al.\egroup }2014b]{chang2014typed}
Kai-Wei Chang, Wen-tau Yih, Bishan Yang, and Christopher Meek.
\newblock 2014b.
\newblock Typed tensor decomposition of knowledge bases for relation
  extraction.
\newblock In {\em Proceedings of the 2014 Conference on Empirical Methods in
  Natural Language Processing (EMNLP)}, pages 1568--1579.

\bibitem[\protect\citename{Chen \bgroup et al.\egroup }2013]{chen-etal-2013}
Yun-Nung Chen, William~Y. Wang, and Alexander~I. Rudnicky.
\newblock 2013.
\newblock {Unsupervised induction and filling of semantic slots for spoken
  dialogue systems using frame-semantic parsing}.
\newblock In {\em 2013 IEEE Workshop on Automatic Speech Recognition and
  Understanding (ASRU)}, pages 120--125. IEEE.

\bibitem[\protect\citename{Chen \bgroup et al.\egroup }2015]{ChenWGR15}
Yun-Nung Chen, William~Yang Wang, Anatole Gershman, and Alexander~I. Rudnicky.
\newblock 2015.
\newblock Matrix factorization with knowledge graph propagation for
  unsupervised spoken language understanding.
\newblock In {\em ACL (1)}, pages 483--494. The Association for Computer
  Linguistics.

\bibitem[\protect\citename{Dong \bgroup et al.\egroup }2014]{dong2014knowledge}
Xin Dong, Evgeniy Gabrilovich, Geremy Heitz, Wilko Horn, Ni~Lao, Kevin Murphy,
  Thomas Strohmann, Shaohua Sun, and Wei Zhang.
\newblock 2014.
\newblock Knowledge vault: A web-scale approach to probabilistic knowledge
  fusion.
\newblock In {\em Proceedings of the 20th ACM SIGKDD international conference
  on Knowledge discovery and data mining}, pages 601--610. ACM.

\bibitem[\protect\citename{Erdos and
  Miettinen}2013]{Erdos:2013:DFB:2505515.2507846}
Dora Erdos and Pauli Miettinen.
\newblock 2013.
\newblock Discovering facts with boolean tensor tucker decomposition.
\newblock In {\em Proceedings of the 22Nd ACM International Conference on
  Information \& Knowledge Management}, CIKM '13, pages 1569--1572, New York,
  NY, USA. ACM.

\bibitem[\protect\citename{Etzioni \bgroup et al.\egroup
  }2011]{etzioni2011open}
Oren Etzioni, Anthony Fader, Janara Christensen, Stephen Soderland, and Mausam
  Mausam.
\newblock 2011.
\newblock Open information extraction: The second generation.
\newblock In {\em IJCAI}, volume~11, pages 3--10.

\bibitem[\protect\citename{Gal{\'{a}}rraga \bgroup et al.\egroup
  }2014]{galarraga2014canonicalizing}
Luis Gal{\'{a}}rraga, Geremy Heitz, Kevin Murphy, and Fabian Suchanek.
\newblock 2014.
\newblock {Canonicalizing Open Knowledge Bases}.
\newblock CIKM.

\bibitem[\protect\citename{Goyal \bgroup et al.\egroup
  }2013]{Goyal_astructured}
Kartik Goyal, Sujay Kumar, Jauhar Huiying, Li~Mrinmaya, Sachan Shashank, and
  Srivastava~Eduard Hovy.
\newblock 2013.
\newblock A structured distributional semantic model: Integrating structure
  with semantics.

\bibitem[\protect\citename{Harshman}1970]{harshman1970fpp}
R.~A. Harshman.
\newblock 1970.
\newblock {Foundations of the PARAFAC procedure: Models and conditions for an"
  explanatory" multi-modal factor analysis}.
\newblock {\em UCLA Working Papers in Phonetics}, 16(1):84.

\bibitem[\protect\citename{Hearst}1992]{Hearst1992}
Marti~A. Hearst.
\newblock 1992.
\newblock Automatic acquisition of hyponyms from large text corpora.
\newblock In {\em In Proceedings of the 14th International Conference on
  Computational Linguistics}, pages 539--545.

\bibitem[\protect\citename{Krompa{\ss} \bgroup et al.\egroup
  }2013]{krompass2013non}
Denis Krompa{\ss}, Maximilian Nickel, Xueyan Jiang, and Volker Tresp.
\newblock 2013.
\newblock Non-negative tensor factorization with rescal.
\newblock {\em Tensor Methods for Machine Learning, ECML workshop}.

\bibitem[\protect\citename{Lee and Seung}2000]{Lee00algorithmsfor}
Daniel~D. Lee and H.~Sebastian Seung.
\newblock 2000.
\newblock Algorithms for non-negative matrix factorization.
\newblock In {\em In NIPS}, pages 556--562. MIT Press.

\bibitem[\protect\citename{Manning \bgroup et al.\egroup
  }2014]{manning-EtAl:2014:P14-5}
Christopher~D. Manning, Mihai Surdeanu, John Bauer, Jenny Finkel, Steven~J.
  Bethard, and David McClosky.
\newblock 2014.
\newblock The {Stanford} {CoreNLP} natural language processing toolkit.
\newblock In {\em Proceedings of 52nd Annual Meeting of the Association for
  Computational Linguistics: System Demonstrations}, pages 55--60.

\bibitem[\protect\citename{Mikolov \bgroup et al.\egroup }2013]{NIPS2013_5021}
Tomas Mikolov, Ilya Sutskever, Kai Chen, Greg~S Corrado, and Jeff Dean.
\newblock 2013.
\newblock Distributed representations of words and phrases and their
  compositionality.
\newblock In C.J.C. Burges, L.~Bottou, M.~Welling, Z.~Ghahramani, and K.Q.
  Weinberger, editors, {\em Advances in Neural Information Processing Systems
  26}, pages 3111--3119. Curran Associates, Inc.

\bibitem[\protect\citename{Mitchell \bgroup et al.\egroup }2015]{NELL-aaai15}
T.~Mitchell, W.~Cohen, E.~Hruschka, P.~Talukdar, J.~Betteridge, A.~Carlson,
  B.~Dalvi, M.~Gardner, B.~Kisiel, J.~Krishnamurthy, N.~Lao, K.~Mazaitis,
  T.~Mohamed, N.~Nakashole, E.~Platanios, A.~Ritter, M.~Samadi, B.~Settles,
  R.~Wang, D.~Wijaya, A.~Gupta, X.~Chen, A.~Saparov, M.~Greaves, and
  J.~Welling.
\newblock 2015.
\newblock Never-ending learning.
\newblock In {\em Proceedings of AAAI}.

\bibitem[\protect\citename{Mohamed \bgroup et al.\egroup
  }2011]{Mohamed11discoveringrelations}
Thahir~P. Mohamed, Estevam~R. Hruschka, Jr., and Tom~M. Mitchell.
\newblock 2011.
\newblock Discovering relations between noun categories.
\newblock In {\em Proceedings of the Conference on Empirical Methods in Natural
  Language Processing}, EMNLP '11, pages 1447--1455, Stroudsburg, PA, USA.
  Association for Computational Linguistics.

\bibitem[\protect\citename{M{\o}rup \bgroup et al.\egroup }2008]{IMM2008-04658}
M.~M{\o}rup, L.~K. Hansen, and S.~M. Arnfred.
\newblock 2008.
\newblock Algorithms for sparse non-negative {TUCKER}.
\newblock {\em Neural Computation}, 20(8):2112--2131, aug.

\bibitem[\protect\citename{Movshovitz-Attias and
  Cohen}2015]{kblda:movshovitzattias-wcohen:2015:ACL}
Dana Movshovitz-Attias and William~W. Cohen.
\newblock 2015.
\newblock Kb-lda: Jointly learning a knowledge base of hierarchy, relations,
  and facts.
\newblock In {\em Proceedings of the 53rd Annual Meeting of the Association for
  Computational Linguistics}. Association for Computational Linguistics.

\bibitem[\protect\citename{Murphy \bgroup et al.\egroup
  }2012]{murphy2012learning}
Brian Murphy, Partha~Pratim Talukdar, and Tom~M Mitchell.
\newblock 2012.
\newblock Learning effective and interpretable semantic models using
  non-negative sparse embedding.
\newblock In {\em COLING}, pages 1933--1950.

\bibitem[\protect\citename{Narita \bgroup et al.\egroup }2012]{usingaux}
Atsuhiro Narita, Kohei Hayashi, Ryota Tomioka, and Hisashi Kashima.
\newblock 2012.
\newblock Tensor factorization using auxiliary information.
\newblock {\em Data Mining and Knowledge Discovery}, 25(2):298--324.

\bibitem[\protect\citename{Nickel \bgroup et al.\egroup
  }2011]{Nickel_athree-way}
Maximilian Nickel, Volker Tresp, and Hans-Peter Kriegel.
\newblock 2011.
\newblock A three-way model for collective learning on multi-relational data.
\newblock In Lise Getoor and Tobias Scheffer, editors, {\em Proceedings of the
  28th International Conference on Machine Learning (ICML-11)}, ICML '11, pages
  809--816, New York, NY, USA, June. ACM.

\bibitem[\protect\citename{Nickel \bgroup et al.\egroup
  }2012]{Nickel:2012:FYS:2187836.2187874}
Maximilian Nickel, Volker Tresp, and Hans-Peter Kriegel.
\newblock 2012.
\newblock Factorizing yago: Scalable machine learning for linked data.
\newblock In {\em Proceedings of the 21st International Conference on World
  Wide Web}, WWW '12, pages 271--280, New York, NY, USA. ACM.

\bibitem[\protect\citename{Riedel \bgroup et al.\egroup }2013]{RiedelYMM13}
Sebastian Riedel, Limin Yao, Andrew McCallum, and Benjamin~M. Marlin.
\newblock 2013.
\newblock Relation extraction with matrix factorization and universal schemas.
\newblock In {\em Human Language Technologies: Conference of the North American
  Chapter of the Association of Computational Linguistics, Proceedings, June
  9-14, 2013, Westin Peachtree Plaza Hotel, Atlanta, Georgia, {USA}}, pages
  74--84.

\bibitem[\protect\citename{Singh \bgroup et al.\egroup }2015]{singh2015towards}
Sameer Singh, Tim Rockt{\"a}schel, and Sebastian Riedel.
\newblock 2015.
\newblock {Towards Combined Matrix and Tensor Factorization for Universal
  Schema Relation Extraction}.
\newblock In {\em NAACL Workshop on Vector Space Modeling for NLP (VSM)}.

\bibitem[\protect\citename{Suchanek \bgroup et al.\egroup
  }2007]{suchanek2007yago}
Fabian~M Suchanek, Gjergji Kasneci, and Gerhard Weikum.
\newblock 2007.
\newblock Yago: a core of semantic knowledge.
\newblock In {\em Proceedings of WWW}.

\bibitem[\protect\citename{Velardi \bgroup et al.\egroup
  }2013]{velardi2013ontolearn}
Paola Velardi, Stefano Faralli, and Roberto Navigli.
\newblock 2013.
\newblock Ontolearn reloaded: A graph-based algorithm for taxonomy induction.
\newblock {\em Computational Linguistics}, 39(3):665--707.

\bibitem[\protect\citename{Wang \bgroup et al.\egroup
  }2015]{conf/kdd/WangCGDKCMS15}
Yichen Wang, Robert Chen, Joydeep Ghosh, Joshua~C. Denny, Abel~N. Kho, You
  Chen, Bradley~A. Malin, and Jimeng Sun.
\newblock 2015.
\newblock Rubik: Knowledge guided tensor factorization and completion for
  health data analytics.
\newblock In Longbing Cao, Chengqi Zhang, Thorsten Joachims, Geoffrey~I. Webb,
  Dragos~D. Margineantu, and Graham Williams, editors, {\em KDD}, pages
  1265--1274. ACM.

\end{thebibliography}
\bibliographystyle{emnlp2016}

\end{document}